\documentclass[aps,prb,twocolumn,superscriptaddress,showpacs]{revtex4}
\usepackage{graphicx}
\usepackage{amssymb}
\usepackage{amsfonts}
\usepackage{bm}
\begin{document}
\title{Ferromagnetic properties of $p$-(Cd,Mn)Te quantum wells:
Interpretation of magneto-optical measurements by Monte
Carlo simulations}

\author{A. Lipi\'nska}
\affiliation{Institute of Physics, Polish Academy of Science,
al.~Lotnik\'ow 32/46, PL 02-668 Warszawa, Poland}

\author{C. Simserides}
\affiliation{Institute of Materials Science, NCSR Demokritos, GR-15310 Athens, Greece}

\author{K. N. Trohidou}
\affiliation{Institute of Materials Science, NCSR Demokritos, GR-15310 Athens, Greece}

\author{M. Goryca}
\affiliation{Institute of Experimental Physics, University of Warsaw, ul. Ho\.za 69, PL 00-681 Warszawa, Poland}

\author{P. Kossacki}
\affiliation{Institute of Experimental Physics, University of Warsaw, ul. Ho\.za 69, PL 00-681 Warszawa, Poland}

\author{A. Majhofer}
\affiliation{Institute of Experimental Physics, University of Warsaw, ul. Ho\.za 69, PL 00-681 Warszawa, Poland}

\author{T. Dietl}
\affiliation{Institute of Physics, Polish Academy of Science,
al.~Lotnik\'ow 32/46, PL 02-668 Warszawa, Poland}
\affiliation{Institute of Theoretical Physics, University of Warsaw, ul. Ho\.za 69, PL 00-681 Warszawa, Poland}

\date{\today}

\begin{abstract}

In order to single out dominant phenomena that account for carrier-controlled magnetism in $p$-$\mathrm{Cd}_{1-x}\mathrm{Mn}_{x}\mathrm{Te}$  quantum wells we have carried out magneto-optical measurements and Monte Carlo simulations of time dependent magnetization. The experimental results show that magnetization relaxation is faster than 20~ns in the paramagnetic state. Decreasing temperature below the Curie temperature $T_{\mathrm{C}}$ results in an increase of the relaxation time but to less than 10~$\mu$s. This fast relaxation may explain why the spontaneous spin splitting of electronic states is not accompanied by the presence of non-zero macroscopic magnetization below $T_{\mathrm{C}}$. Our Monte Carlo results reproduce the relative change of the relaxation time on decreasing temperature. At the same time, the numerical calculations demonstrate that antiferromagnetic spin-spin interactions, which compete with the hole-mediated long-range ferromagnetic coupling, play an important role in magnetization relaxation of the system. We find, in particular, that magnetization dynamics is largely accelerated by the presence of antiferromagnetic couplings to the Mn spins located outside the region, where the holes reside. This suggests that macroscopic spontaneous magnetization should be observable if the thickness of the layer containing localized spins will be smaller than the extension of the hole wave function. Furthermore, we study how a spin-independent part of the Mn potential affects $T_{\mathrm{C}}$. Our findings show that the alloy disorder potential tends to reduce $T_{\mathrm{C}}$, the effect being particularly strong for the attractive potential that leads to hole localization.
\end{abstract}


\maketitle

\section{Introduction}
\label{sec:introduction}
Modulation-doped $p$-type (Cd,Mn)Te/(Cd,Mg,Zn)Te quantum wells (QWs)
remain a unique medium allowing to probe carrier-induced Ising-like ferromagnetism in the two-dimensional (2D) case,\cite{Haury:1997,Kossacki:2000,Kossacki:2002,Kossacki:2006,Maslana:2006} as in this system the mean free path is longer than the QW width.
The presence of ferromagnetism was revealed
by the observation of spontaneous splitting of the photoluminescence (PL) spectrum,
which indicated that local spin ordering is stable, below the Curie temperature $T_{\mathrm{C}}$,
for times longer than the exciton lifetime.\cite{Kossacki:2000,Boukari:2002}
Theoretical analyzes have shown that the magnitude of $T_{\mathrm{C}}$ as a function of the Mn concentration $x$ and the hole areal density $p$
can be quantitatively described provided that in addition to carrier-mediated ferromagnetic spin-spin couplings,
the presence of competing short-range antiferromagnetic superexchange interactions as well as
hole correlation effects are taken into account.\cite{Dietl:1997,Haury:1997,Dietl:1999,Boukari:2002,Kechrakos:2005}
These experiments and their quantitative interpretations were carried out prior to later theoretical studies.\cite{Lee:2000,Brey:2000,Priour:2005}

A surprising result of both optical and magnetic measurements is the absence of hysteresis loops and,
hence, of macroscopic spontaneous magnetization below $T_{\mathrm{C}}$.\cite{Kossacki:2000,Kossacki:2002}
This finding questions our understanding of the actual ground state of the system.
It has been suggested, in particular, that the formation of spin density waves,
driven by the $q$-dependent carrier susceptibility, may account for the absence of spontaneous magnetization.\cite{Kossacki:2000}
This possibility has recently been considered in the context of
theoretical search for carrier-mediated nuclear magnetism in 2D systems.\cite{Simon:2008}

In order to determine the importance of various phenomena that control magnetism in such reduced-dimensionality magnetically disordered systems
we have carried out time-resolved magneto-optical measurements.
Furthermore, with the goal to obtain information on mechanisms controlling spin dynamics,
we have extended our previous\cite{Kechrakos:2005} Monte Carlo (MC) simulations of magnetization in
$p-\mathrm{Cd}_{1-x}\mathrm{Mn}_{x}\mathrm{Te}$ QW.
The experimental findings provide a dependence of magnetization relaxation on temperature and carrier density.
Above $T_{\mathrm{C}}$ the relaxation time is found to be shorter than 20~ns,
while below $T_{\mathrm{C}}$ magnetization persists up to a few microseconds,
a time scale consistent with the absence of spontaneous magnetization in the static measurements.
Interestingly, we have obtained a similar relative prolongation of the relaxation time below $T_{\mathrm{C}}$
from MC simulations based on the Metropolis algorithm and on the determination of the one-particle hole eigenfunctions at each MC sweep.
The simulations explain also the rate of increase of the relaxation time with the carrier density.
We conclude that the absence of magnetic hysteresis in this 2D Ising system can be explained without taking carrier-carrier correlation into account.

At the same time, by analyzing results of our Monte Carlo simulations,
we are able to show that short range spin-spin antiferromagnetic (AFM) interactions  play a crucial role in accelerating magnetization dynamics.
This corroborates the outcome of the previous results showing a narrowing of the hysteresis loop by AFM interactions.\cite{Kechrakos:2005}
However, we realize now, based on much more extensive simulations, that the effect of AFM interactions becomes much reduced
if the thickness of the layer containing Mn spins is taken to be smaller than the extent of the hole wave function.
This finding, from the one hand, substantiates theoretical considerations of Boudinet and Bastard,\cite{Boudinet:1993}
suggesting that magnetization relaxation of bound magnetic polarons in $p$-type CdTe/(Cd,Mn)Te QWs occurs
owing to the AFM coupling to the Mn spins located outside the relevant Bohr radius.
On the other, this result implies that macroscopic spontaneous magnetization and magnetic hysteresis should be recovered in quantum wells,
in which the thickness of the Mn layer would be smaller than the region visited by the holes.

Furthermore, we study how a spin-independent part of the potential introduced by Mn impurities affects $T_{\mathrm{C}}$.
Our simulations show that alloy disorder tends to reduce $T_{\mathrm{C}}$.
The effect is particularly dramatic for the attractive alloy potential  which, if sufficiently large, leads to a strong hole localization.
This result substantiates the notion that delocalized or weakly localized carriers are necessary
to generate a sizable ferromagnetic coupling between diluted localized spins.

This article is organized as follows.
In Sec.~\ref{sec:experiments} we present experimental findings of magneto-optical studies on  magnetization dynamics,
which have motivated the theoretical effort.
The theoretical model
that describes the carrier-induced ferromagnetism in
diluted magnetic semiconductor (DMS) quantum wells is introduced
in Sec.~\ref{sec:theory}.
In Sec.~\ref{sec:tc} we present MC investigations of
the dependence of the critical temperature on the carrier density and spin-independent alloy potential. Section~\ref{sec:dyn-dom} contains results of our MC studies of spin dynamics and of spin ordering, which serve to interpret the findings of magneto-optical measurements.
Specifically,
subsection~\ref{subsec:dyn} is devoted to magnetization dynamics, while
subsection~\ref{subsec:dom} contains a discussion of possible domain sizes.
Our conclusions are summarized in Sec.~\ref{sec:conclusion}.

\section{Experimental results}
\label{sec:experiments}
The experimental results have been obtained for samples grown in Grenoble by molecular beam epitaxy on $\mathrm{Cd}_{0.88}\mathrm{Zn}_{0.12}\mathrm{Te}$ (001) substrates.\cite{Haury:1997,Kossacki:2004}
The modulation-doped $p$-type structures
contain a single 8~nm wide  $\mathrm{Cd}_{1-x}\mathrm{Mn}_{x}\mathrm{Te}$ QW embedded by
$\mathrm{Cd}_{1-y-z}\mathrm{Mg}_{y}\mathrm{Zn}_{z}\mathrm{Te}$ barriers, in which the Mg
content ($y$ = 0.25-0.28) results in a sizable valence band offset, while
the presence of Zn ($z$ = 0.07--0.08) ensures a good lattice match to the substrate.
The front barrier is doped by nitrogen acceptors.
The distance between the QW and the doping layer is 20~nm, which results in the
hole density up to $3\times10^{11} \mathrm {cm^{-2}}$.\cite{Kossacki:2004} These holes occupy the ground state subband whose scattering broadening is much smaller than the distance to the next subband, making the system to be truly 2D.

According to the well established procedure, magnetic properties of the QW are probed by
magnetospectroscopy.\cite{Haury:1997,Kossacki:2000,Boukari:2002}
The photoluminescence (PL) is excited by a cw laser with the photon energies
below the absorption edge of the barrier material.
A typical PL spectrum of a paramagnetic $p$-type QW consists of a single line
related to the radiative recombination of charged excitons ($\mathrm{X^{+}}$).
Depending on its polarization chirality, the emission line exhibits a downward or upward shift in the magnetic field.
Owing to a strong $sp$-$d$ exchange interaction specific to DMSs, this Zeeman shift is giant (typically over 20~meV in 1~T), and
its magnitude is proportional to the local magnetization of the Mn spins.
Importantly, a combination of strain, confinement, and spin-orbit interaction makes that
the giant shift vanishes virtually entirely for the in-plane magnetic field.
Below a critical temperature, which we identify as the Curie temperature $T_{\mathrm{C}}$,
a spontaneous splitting of the PL line into two components is observed in the absence of an external magnetic field.
The energy distance between these two lines, or the downward shift of the lower line,
provides information on an average value of spontaneous magnetization in the regions visited by the holes at given temperature.

Surprisingly, the emitted light at the wavelength corresponding to either of these two lines remains unpolarized,
even after ramping the field down to zero below $T_{\mathrm{C}}$.
A magnetic field of the order of 20~mT, much higher than the demagnetization field for this diluted alloy,
is necessary to achieve a full circular polarization of the lines at $T \approx 0.5T_{\mathrm{C}}$.\cite{Kossacki:2000}
This indicates the absence of a macroscopic spontaneous magnetization at these low temperatures,
the conclusion confirmed by a direct magnetization measurements.\cite{Sawicki:2002}
Moreover, even focusing detection on the spot with diameter below 1~$\mu$m does not result in the circularly polarized emission.
This could be explained either by a small size of the relevant magnetic domains or by
fast fluctuations of the magnetization direction.
Alternatively, a spin-density wave could be considered as the relevant ground state of this reduced dimensionality correlated system.
However, a selective excitation of PL by circularly polarized light leads to circularly polarized PL.\cite{Kossacki:2002}
This indicates that local spin ordering is stable for times longer than the exciton lifetime, estimated to be in the 100~ps range.

Some of us have recently developed a technique for studies of magnetization dynamics by means of time resolved PL measurements
performed after a short pulse of the magnetic field.\cite{Kossacki:2006,Maslana:2006}
Pulses of the magnetic field are produced by a magnetic coil mounted at the surface of the sample.
The illumination with a laser beam and the collection of the PL signal are performed
along the axis of the coil (Faraday configuration).
The coil diameter of 0.5~mm results in a small value of inductance and allows to obtain short rise and fall times of about 10~ns.
A 2~A current produces a magnetic field of about 40~mT.
The time evolution of the PL during and after the pulse is probed with resolution down to 10~ns.
Magnetization dynamics is studied by analyzing the difference in PL intensities
for two circular polarizations after the pulse of the magnetic field.
Results of measurements shown here have been obtained at different temperatures below and above
$T_{\mathrm{C}}$, which is 2.5~K for the studied $\mathrm{Cd}_{0.96}\mathrm{Mn}_{0.04}\mathrm{Te}$ QW.

Above $T_{\mathrm{C}}$, in the paramagnetic state, a single PL line is observed at zero field, and
the signal induced by the magnetic pulse is related to Zeeman splitting of this line.
Its relaxation time is found to be shorter than 20~ns.
Such a value is just to be expected for (Cd,Mn)Te containing 4\% of Mn.\cite{Scalbert:1988, Dietl:1995}

\begin{figure}
\includegraphics[scale=1.0]{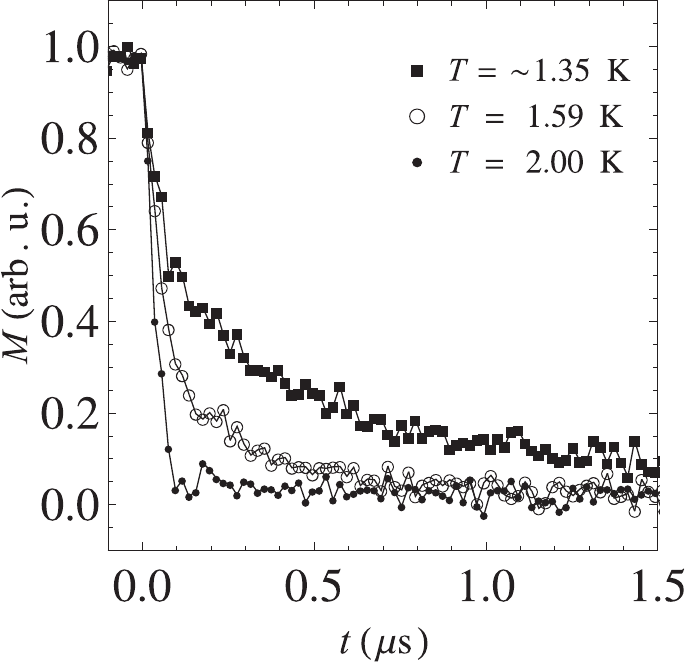}
\caption{Experimental: Temporal decay of magnetization in the Cd$_{0.96}$Mn$_{0.04}$Te quantum well at various temperatures
determined as a difference in photoluminescence intensities collected for two circular polarizations
after a pulse of the magnetic field. The Curie temperature is 2.5~K.}
\label{fig:corfun1}
\end{figure}

Typical temporal profiles for three different temperatures below $T_{\mathrm{C}} = 2.5$~K are presented in Fig.~\ref{fig:corfun1}.
The relaxation time becomes much slower when decreasing temperature (see Fig.~\ref{fig:corfun1}).
Magnetization persists up to a few microseconds when $T \approx 0.5T_{\mathrm{C}}$.
However even for the lowest temperatures, the relaxation time is shorter than 10~$\mu$s, and
under all experimental conditions the remanent magnetization vanishes totally after 20~$\mu$s.
This time scale is much shorter than an acquisition time of standard optical measurements,
which accounts for unmeasurably small values of the coercive field observed experimentally.

\begin{figure}
\includegraphics[scale=1.0]{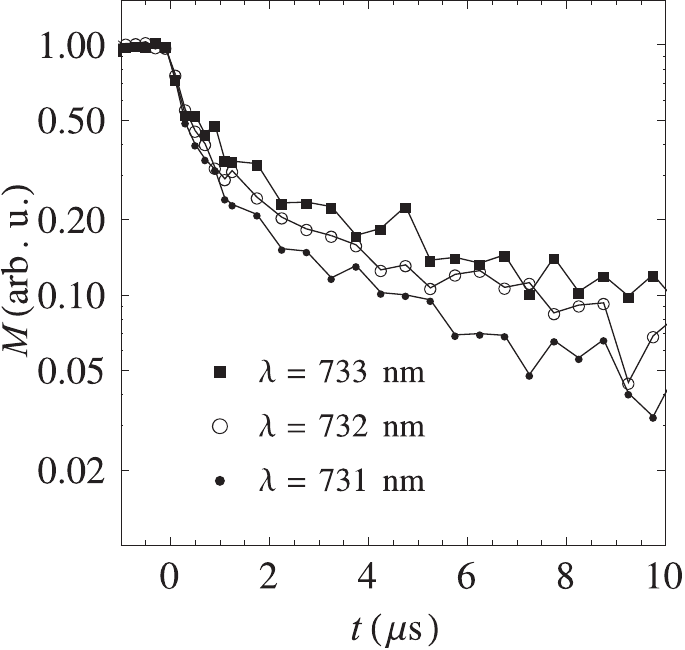}
\caption{Experimental: Temporal decay of magnetization in the Cd$_{0.96}$Mn$_{0.04}$Te quantum well
determined as a difference in photoluminescence intensities for two circular polarizations
collected at various wavelengths after a pulse of the magnetic field. The Curie temperature is 2.5~K.} \label{fig:corfun2}
\end{figure}

It is interesting to note that the decays are not mono-exponential.
The relaxation begins with a rapid decay, faster than 100~ns, but
it also contains a slower component, in the microsecond range.
One could expect that regions of higher hole density have higher local magnetization
resulting in higher barriers for the magnetization reversal.
In order to examine this possibility we carried out relaxation measurements for a constant temperature (below $T_{\mathrm{C}}$), and
different detection wavelength which was tuned over the low energy component of the PL line.
As mentioned above, this line shows a red shift when decreasing temperature below $T_{\mathrm{C}}$.
This shift is proportional to Mn magnetization, whose magnitude is primarily determined by the hole density.\cite{Boukari:2002,Kossacki:2004}
Due to spatial fluctuations in the hole density,\cite{Maslana:2006} the line is significantly broadened.
The low energy side of the line comes from the regions containing higher hole densities,
while the high energy side corresponds to locations visited by fewer holes.
This provides an opportunity to trace the relaxation time as a function of the hole concentration.
The obtained temporal magnetization profiles are presented in Fig.~\ref{fig:corfun2}.
It is  seen that the contribution of the slow component increases for the regions containing a high hole density.
This finding as well as the observed temperature variation of the magnetization decay
time will be compared to results of numerical simulations in the next section.

\section{Theory and simulation procedure}
\label{sec:theory}
To understand the experimental results described in Sec.~\ref{sec:experiments} we carried out Monte Carlo simulations for
the subsystem of Mn spins which are at the same time subject to short-range antiferromagnetic and
long-range hole-mediated ferromagnetic spin-spin interactions.
According to our results, the presence of these two kinds of competing interactions accounts for
a non-standard character of ferromagnetism in the studied system.
Since we are mainly interested in modulation-doped $p$-Cd$_{1-x}$Mn$_{x}$Te quantum wells,
all our results are obtained using the parameter values characteristic for this system, but
the methodology we use may as well be applied to other 2D $p$-type II-VI DMS structures.

The antiferromagnetic interactions originate from  superexchange, the mechanism specific to undoped charge transfer insulators, such as (Cd,Mn)Te.
The character and magnitude of these interactions are known from previous extensive experimental studies of magnetic properties of (Cd,Mn)Te.
In order to find out how the presence of holes affects the system we compute, in the spirit of the adiabatic approximation,
the total energy of the holes at a given configuration of the Mn spins.
In this way our approach encompasses automatically the description of carrier-mediated exchange interactions within
either $p$-$d$ Zener model or  Ruderman-Kittel-Kasuya-Yosida theory.
The carrier total energy is evaluated neglecting hole correlations,
the approximation that leads to an underestimation of the ferromagnetic coupling.\cite{Dietl:1997}
Accordingly, our computed Curie temperatures are expected to be systematically lower than those observed experimentally.
Furthermore, in accord with the effective mass theory, and owing to a relatively small disorder in our structures,
we assume that the in-plane and the perpendicular hole motions can be factorized,
i.e., that the total envelope function assumes the form
$\Psi_{\sigma}(\bm{R}) = \psi_{\sigma}(\bm{r})\varphi(z)$, where $\bm{R} = (\bm{r},z)$
with in-plane and perpendicular positions described by $\bm{r} = (x,y)$ and $z$ respectively;
$\varphi(z)$ is the wave function of a hole residing in the ground state subband of the QW in question.
For the biaxial strain under consideration, the relevant subband is of the heavy hole origin,
so that the spin-orbit interaction fixes the hole spin in the direction of the growth axis.
Furthermore, since the heavy holes penetrate only weakly the barriers,
$\varphi(z)$ in the form corresponding to an infinite QW is adequate for our computations.

Within the above model and in the absence of an external magnetic field, the Hamiltonian  $\mathcal{H}$ of our system takes the form,

\begin{equation}\label{htotal}
\mathcal{H} = \frac{p^2}{2m^*} + \mathcal{H}_{pd} + \mathcal{H}_{dd}.
\end{equation}

\noindent
This Hamiltonian describes noninteracting carriers in the heavy hole parabolic valence subband
for which the in-plane effective mass is $m^{*} = 0.25m_{0}$ for the QW in question.\cite{Fishman:1995}
Since the coupling between the holes and the Mn ions is weak in (Cd,Mn)Te,\cite{Dietl:2008}
we can consider the $p$-$d$ interaction in a simple contact form,

\begin{equation}\label{hpd}
\mathcal{H}_{pd} = \sum_{i} \frac{1}{3} \beta j_{z} S_{zi}
\delta(\bm{r}- \bm{r}_{i}) \left|\varphi(z_{i})\right|^{2},
\end{equation}

\noindent
where the hole spin $j = 3/2$, $j_{z} = \pm $ 3/2,
is coupled to the Mn ions randomly distributed over the fcc lattice sites $\bm{r}_{i}$.
We treat Mn spins $\bm{S}_{i}$ as classical vectors with $S = 5/2$.
While this approximation is certainly qualitatively valid for the large spin in question, quantitatively -- according to the mean-field theory --
it leads to an underestimation of the $T_{\mathrm{C}}$ value by $(S+1)/S$.
For the $p$-$d$ exchange integral we take $\beta =  -5.96 \times 10^{-2}$~eV~nm$^3$, which corresponds
to the exchange energy $\beta N_0 = -0.88$~eV, where $N_0$ is the cation concentration.\cite{Gaj:1979}

We have also tested how our results are affected by the spin-independent part of the potentials introduced by Mn ions.
To take this alloy disorder into account we add to Eq.~\ref{hpd} a term

\begin{equation}\label{hAD}
\mathcal{H}_{\mathrm{AD}} = \delta V \sum_{i} \delta(\bm{r}- \bm{r}_{i})\left|\varphi(z_{i})\right|^{2}
\end{equation}

\noindent
with  $\delta V = 3.93 \times 10^{-2}$~eV~nm$^3$
which corresponds to the valence band offset $W N_0 = -\delta V N_0 = -0.58$~eV in (Cd,Mn)Te.\cite{Gaj:1994,Wojtowicz:1996}
It is known that the effect of the Mn isoelectronic impurities on the carrier wave function is governed by the ratio of the total Mn potential $U$ to its critical value $U_c <0$ at which a bound state starts to form,\cite{Dietl:2008}

\begin{equation}\label{uperuc}
U/U_c = 6m^*[W - (S + 1)\beta/2]/(\pi^3\hbar^2b),
\end{equation}

\noindent
where $b$ is the potential radius. In the case under consideration, the hole spin is fixed along $z$-axis,
and Mn polarization $\langle S_z \rangle$ is typically below 10\%.
Accordingly, for $S+1 \rightarrow \langle S_z\rangle$ the total Mn potential is expected to be effectively repulsive for the holes in (Cd,Mn)Te, $U/U_c <0$. Interestingly, in many important systems, such as oxides and nitrides, the spin-independent part of the magnetic impurity potential is actually attractive.\cite{Dietl:2008} Accordingly, we have also performed some simulations of magnetization for $\delta V = - 3.93 \times 10^{-2}$~eV~nm$^3$
which corresponds to the valence band offset $W N_0 = -\delta V N_0 = +0.58$~eV.
However, it should be noted that by using the $\delta$-like potentials, which corresponds to $b\rightarrow 0$, we actually overestimate $U/U_c$ and
determine an upper limit of the effect of the alloy potential upon $T_{\mathrm{C}}$.
With this in mind, we have carried out all MC {\em dynamics} studies presented in Sec.~\ref{sec:dyn-dom} for $\delta V = 0$.

The short-range intrinsic antiferromagnetic (AFM) interaction between the Mn spins is given by,
\begin{equation}\label{hdd}
\mathcal{H}_{dd} = -2k_{B} \sum_{ij}  J_{ij} \bm{S}_{i} \cdot \bm{S}_{j},
\end{equation}
\noindent
with $J_{ij} = -6.3, -1.9, -0.4$~K for the nearest, next nearest, and next next nearest neighbors, respectively.\cite{Shapira:2002}
Our MC simulations confirm that these values of $J_{ij}$ reproduce correctly the temperature dependence of the spin susceptibility
in the absence of holes but we note that other sets of $J_{ij}$ values have also been proposed in the literature.\cite{Hennion:2002}

The above approach disregards the presence of long-range dipole-dipole interactions between Mn spins.
Both theoretical evaluations and MC simulations we have carried out
show that the dipole-diploe interactions can be safely ignored in the case under consideration.

In order to determine hole eigenfunctions and eigenvalues for a given configuration of Mn spins
we assume periodic boundary conditions in the QW atomic layers and diagonalize $\mathcal{H}$
in a plane-wave basis with 2D wave vectors truncated
at a radius $k_c = 5.52 (2\pi/L)$ or $k_c = 8.10 (2\pi/L)$.\cite{Schliemann:2001}
This part of the 2D $k$-space corresponding to 97 or 213 $k$-states, respectively,
is sufficient to ensure the convergence for all hole density values considered here.
The energy of the holes is determined by summing up the lowest eigenvalues corresponding to a given number of holes $N_\mathrm{h}$.
This procedure assumes that the hole gas is degenerate, i.e., neglects thermal broadening of the hole distribution,
the effect considered previously,\cite{Boukari:2002} and found to be small.
In order to keep $k$-space shells filled up,  we have considered $N_\mathrm{h} = $ 1, 5, 9, 13, 21, 25, 29, 37, 45, 49, and 57,
which correspond to values of hole concentrations $p$ up to $1.11 \times 10^{11}$~cm$^{-2}$ for the chosen size of the simulation box.

In our MC simulations we employ the Metropolis algorithm in which the hole eigenfunctions and eigenvalues are updated at each MC sweep
(in one MC sweep all Mn spins are rotated).
This procedure if applied after each single spin rotation would be computationally excessively time consuming and,
therefore, we followed the idea of the so-called perturbative Monte Carlo method.\cite{Troung:1996,Troung:1997,Kennett:2002}
We typically keep 2000 initial MC sweeps (to thermalize the system),
followed by $10^{4}$-$10^{5}$ MC sweeps used for a further analysis.
Most of the numerical results presented in this paper have been obtained for the magnetic field put to zero.
However, our software allows as well simulations in non-zero magnetic fields.
We have occasionally used this opportunity to obtain some of the initial configurations.

For the computations presented here we adopt the simulation box of dimensions $L \times L \times L_{W}$, where $L \leq 350a_o$ and $L_{W} = 8a_o$,
with the fcc lattice constant $a_o = 0.647$~nm in our case.
Hence, the quantum well width is $L_W = 5.2$~nm and the area $L^2 = $(226~nm)$^2$.
We note that a typical QW width of experimental samples is somewhat grater, $L_W^{exp} = 8$~nm.
However, the adoption of a smaller value of $L_W$ allows us to treat systems with a larger area and,
at the same time, since within the mean-field theory\cite{Dietl:1997} $T_{\mathrm{C}} \propto 1/L_W$,
results in a partial compensation of a systematic error stemming from treating the spins classically,
$L_W^{exp}S/[L_W(S+1)] = 1.1$.
We randomly occupy 4\% of cations sites, which corresponds to 156800 Mn spins.
In view of the required accuracy, this large system size makes averaging over the Mn distribution
and extrapolation to an infinite system size unnecessary.
The number of Monte Carlo sweeps needed for the convergence is of the order of $10^4$ to $10^5$.
Hence, we have to use a (pseudo)random number generator with a sufficiently long period.
After testing various generators, the Mersenne Twister developed by Matsumoto and Nishimura has been selected,
because of its huge period and its very high order of dimensional equidistribution.\cite{matsumoto_nishimura}

\section{Curie temperature}
\label{sec:tc}
The critical temperature $T_{\mathrm{C}}$ has been determined in the standard way:
by MC simulations we have calculated the temperature dependence of the spin projections
and the spin susceptibility for both hole and magnetic ion spins.
To denote spins we use the notation $\sigma = s$ (for holes) and $\sigma = S$ (for Mn ions).
Similarly, $N = N_\mathrm{h}$ (for the number of holes) and $N = N_\mathrm{Mn}$ (for the number of Mn ions).
At each MC sweep, the spin projections per hole or per Mn ion are given by:

\begin{equation}\label{sigma-per-particle}
\overline{\sigma_j} = \frac{\sum_{i=1}^N \sigma_{ij}}{N}, \hspace{0.5cm} j=x,y,z.
\end{equation}

\begin{figure*}
\includegraphics[scale=0.7]{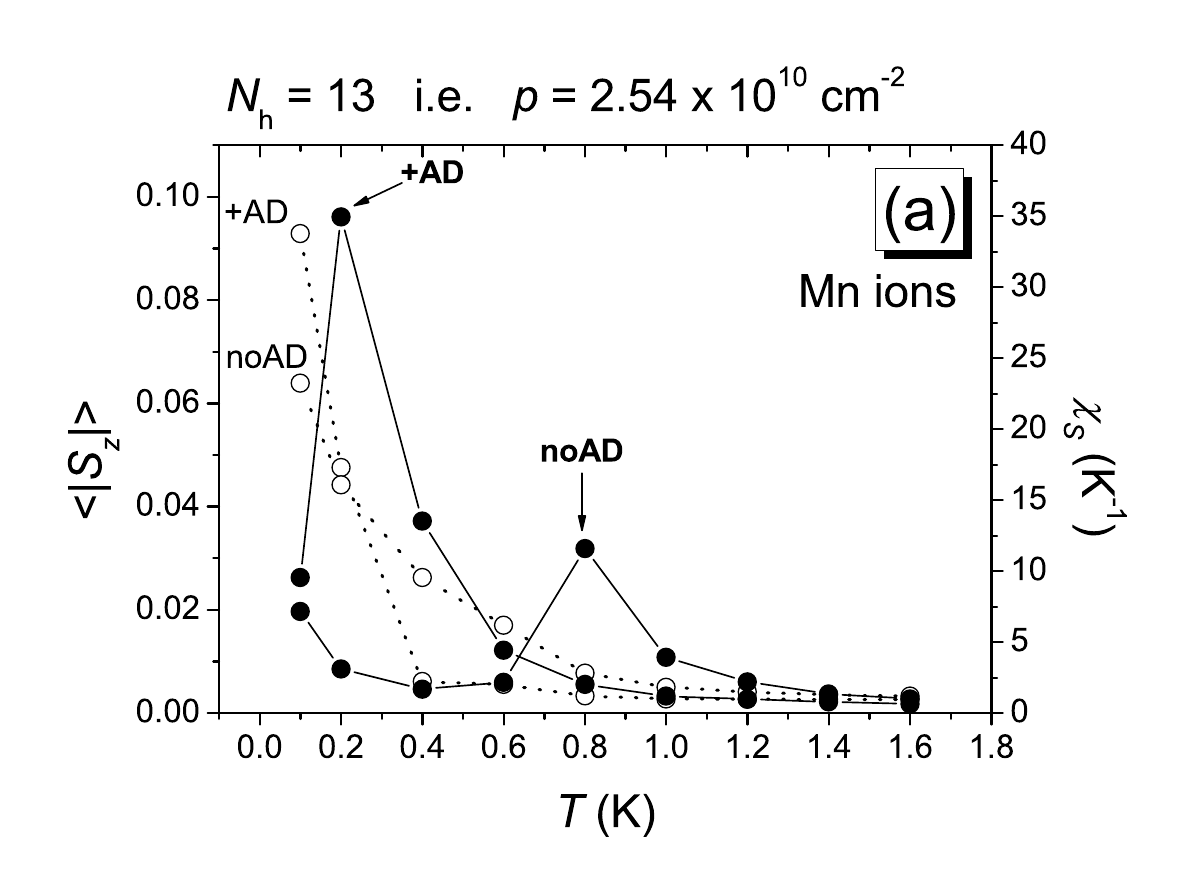}
\includegraphics[scale=0.7]{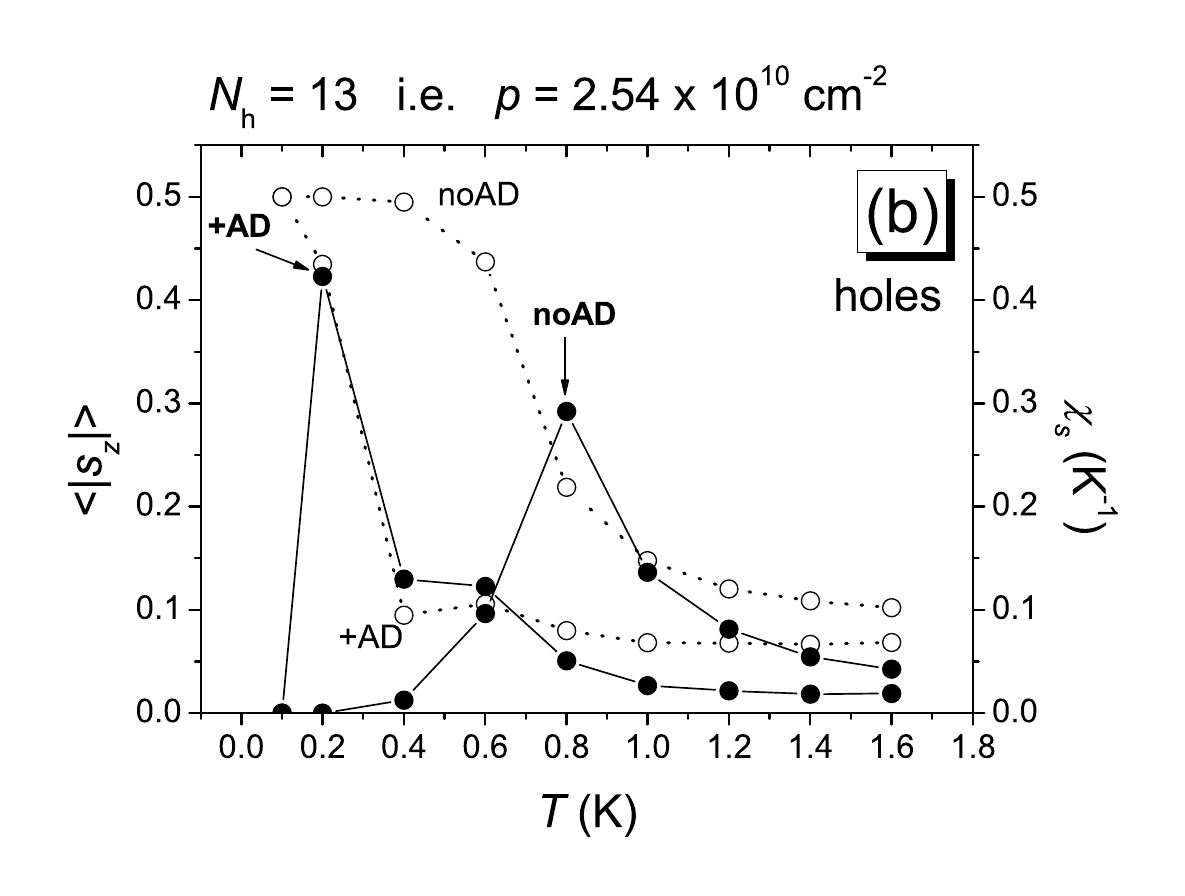}
\includegraphics[scale=0.7]{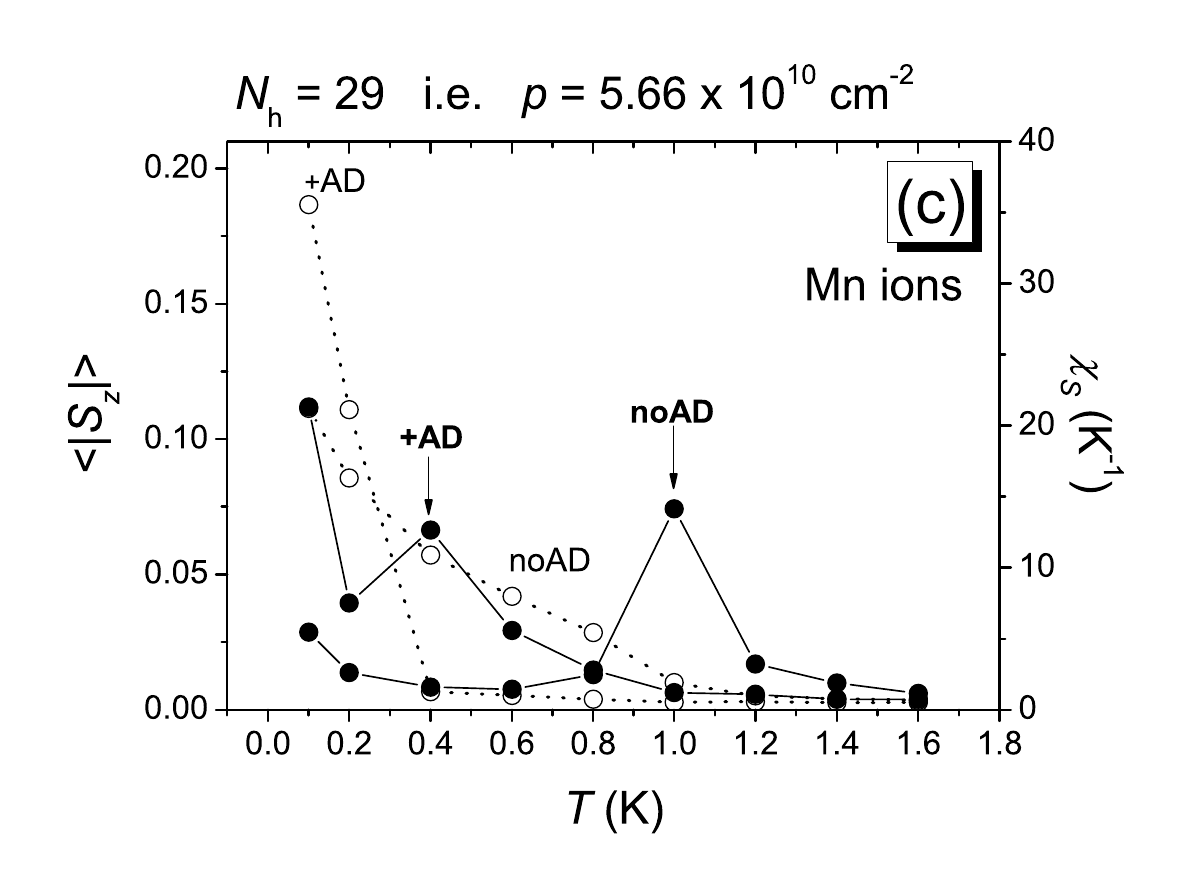}
\includegraphics[scale=0.7]{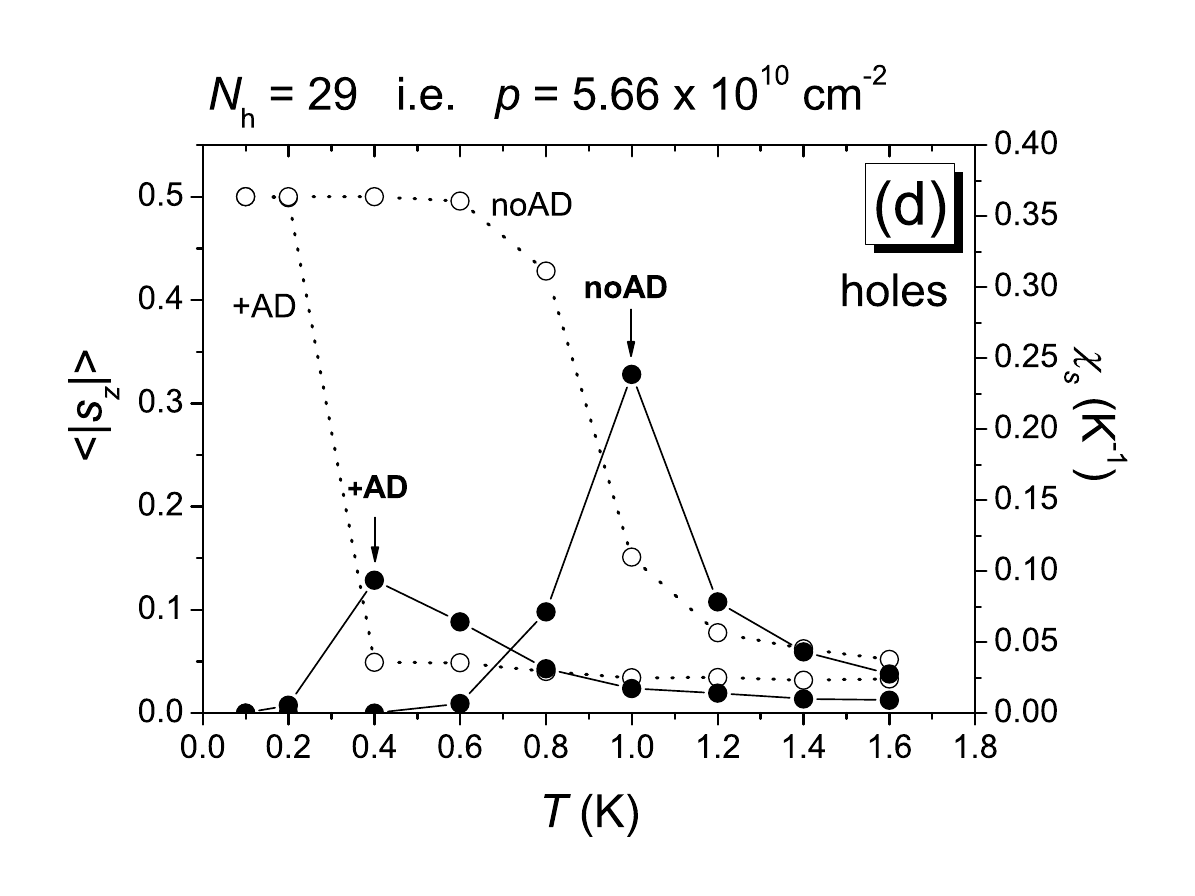}
\caption{Monte Carlo simulations:
the $z$-component of the absolute spin projections
$\langle |S_z| \rangle$ and $\langle |s_z| \rangle$ (open symbols - dotted lines; left scale) and
the spin susceptibilities $\chi_S$ and $\chi_s$ (full symbols - solid lines; right scale)
for the Mn ions (a,c) and for the holes (b,d),
in the absence of alloy disorder (noAD, $\delta V = 0$), and for the repulsive (+AD, $\delta V > 0$) alloy disorder potential.
A maximum in the temperature dependence of spin susceptibility is identified as the Curie temperature below which the holes induce a long range ferromagnetic order.
Here the number of holes $N_\mathrm{h} = 13$ (a,b) and $29$ (c,d), i.e., $p = 2.54 \times 10^{10}$ and $5.66 \times 10^{10}$~cm$^{-2}$, respectively.}
\label{fig:Nh13_29_of_T_+AD_noAD}
\end{figure*}

\begin{figure*}
\includegraphics[scale=0.7]{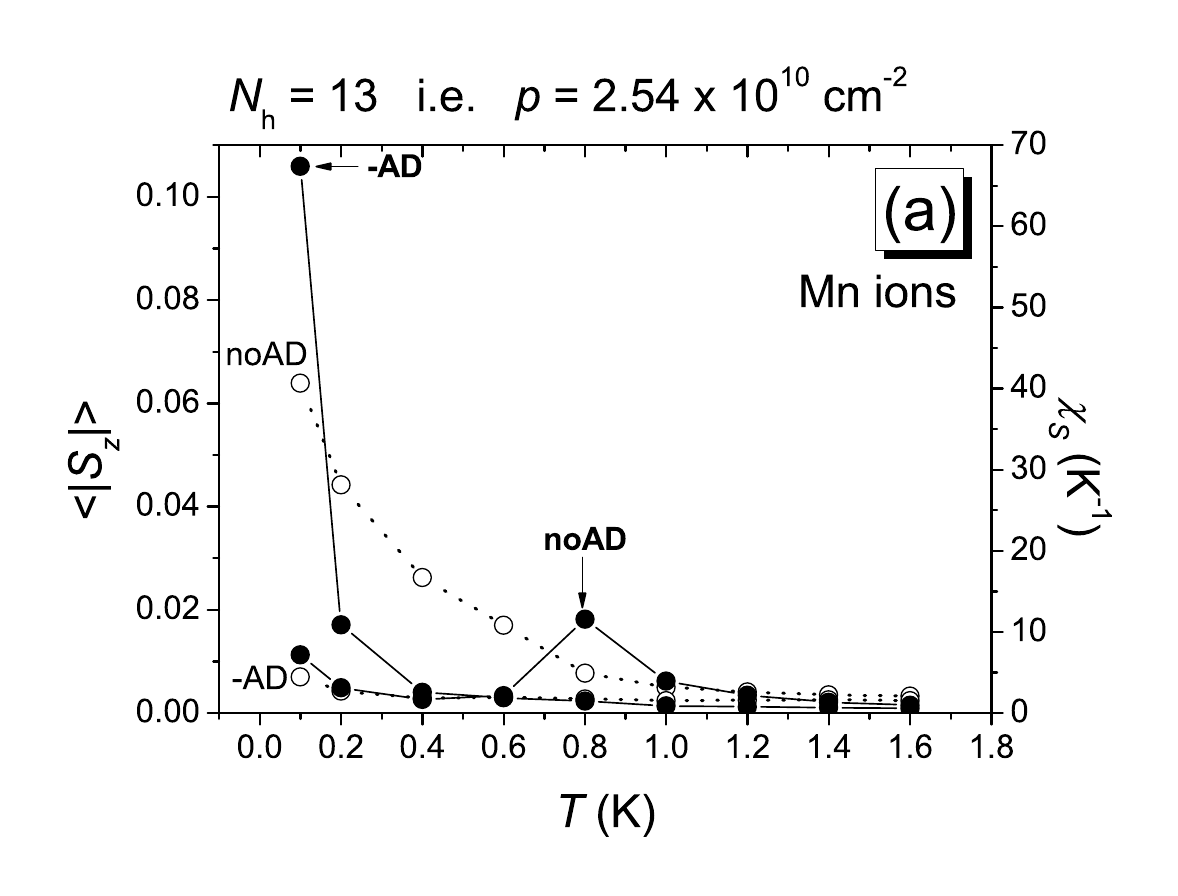}
\includegraphics[scale=0.7]{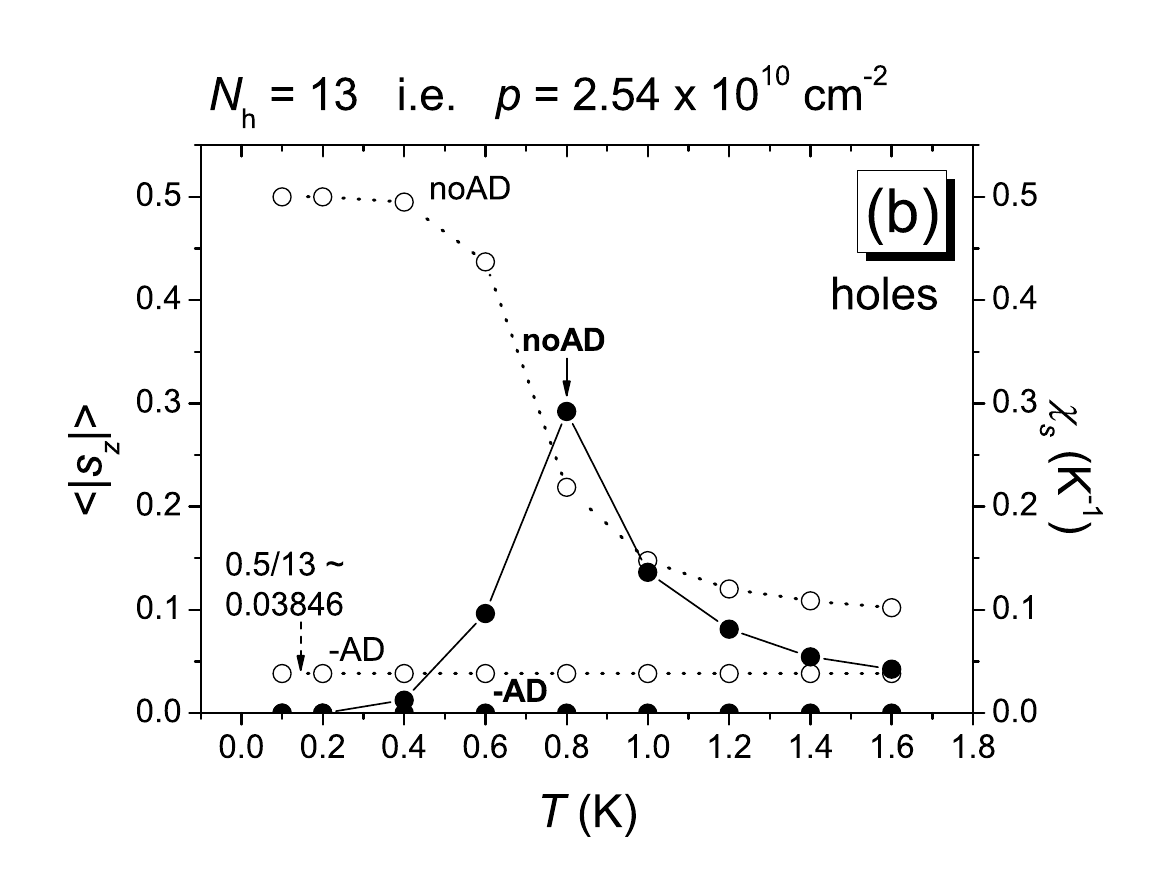}
\includegraphics[scale=0.7]{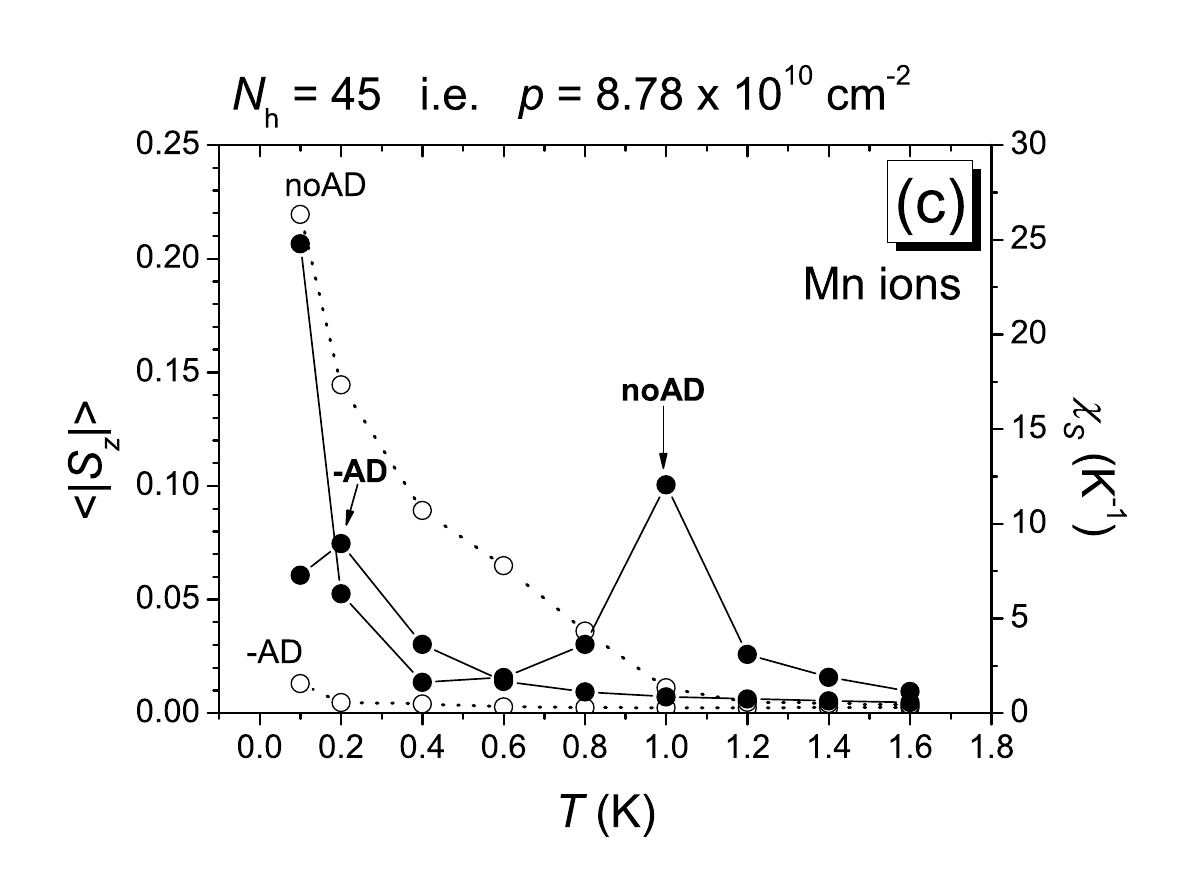}
\includegraphics[scale=0.7]{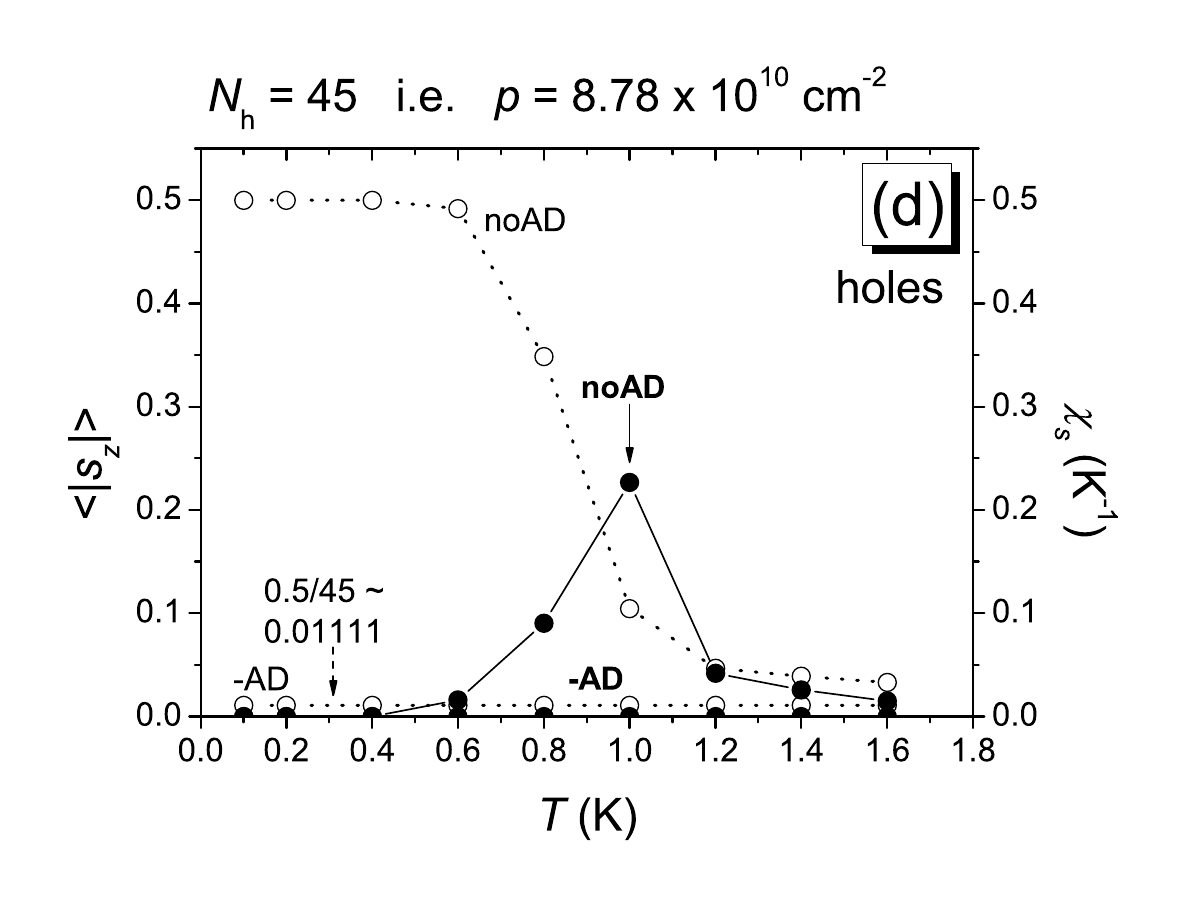}
\caption{Monte Carlo simulations:
the $z$-component of the absolute spin projections
$\langle |S_z| \rangle$ and $\langle |s_z| \rangle$ (open symbols - dotted lines; left scale) and
the spin susceptibilities $\chi_S$ and $\chi_s$ (full symbols - solid lines; right scale)
for the Mn ions (a,c) and for the holes (b,d),
in the absence of alloy disorder (noAD, $\delta V = 0$), and for the attractive (-AD, $\delta V < 0$) alloy disorder potential.
A maximum in the temperature dependence of spin susceptibility is identified as the Curie temperature below which the holes induce a long range ferromagnetic order.
Here the number of holes $N_\mathrm{h} = 13$ (a,b) and $45$ (c,d), i.e., $p = 2.54 \times 10^{10}$ and $ 8.78 \times 10^{10}$ cm$^{-2}$, respectively.}
\label{fig:Nh13_45_of_T_noAD_-AD}
\end{figure*}

\noindent
We symbolize statistical averages of the above quantities by $\langle ... \rangle$.\cite{Newman:1999}
In other words, e.g.

\begin{equation}\label{sigma-st-av}
\langle \sigma_j \rangle = \frac{\sum_{n=1}^{nsum}\overline{\sigma_j}}{nsum}
\end{equation}
\noindent
and
\begin{equation}\label{sigma-st-av-abs}
\langle |\sigma_j| \rangle = \frac{\sum_{n=1}^{nsum}
|\overline{\sigma_j}|}{nsum},
\end{equation}
\noindent
where $n$ denotes successive MC sweeps used for the statistical average and
$nsum$ denotes the total number of MC sweeps used for the statistical average.
The spin susceptibilities per spin are defined in the following manner:\cite{Newman:1999}

\begin{equation}\label{suscept}
\chi_{\sigma_j} = \frac{1}{T}[\langle \sigma_j^2 \rangle - \langle \sigma_j \rangle^2].
\end{equation}
\noindent
We associate the temperature at which the magnitude of spin susceptibility reaches a maximum
with the Curie temperature $T_{\mathrm{C}}$ of the system.
Usually, the MC computations have been performed with the step of 0.2~K.
Therefore, an uncertainty in $T_{\mathrm{C}}$ values is usually at least $\pm 0.1$~K.
Some examples of the temperature dependence of  the spin projections and of the spin susceptibilities
are presented in Fig.~\ref{fig:Nh13_29_of_T_+AD_noAD}.
Specifically, we have plotted $\langle |S_z| \rangle$ and $\chi_S$ for the Mn ions
(Figs.~\ref{fig:Nh13_29_of_T_+AD_noAD}(a) and \ref{fig:Nh13_29_of_T_+AD_noAD}(c))
as well as
$\langle |s_\mathrm{h}| \rangle$ and $\chi_\mathrm{h}$ for the holes
(Fig.~\ref{fig:Nh13_29_of_T_+AD_noAD}(b) and \ref{fig:Nh13_29_of_T_+AD_noAD}(d)),
for the hole numbers $N_\mathrm{h} =  13$ and $29$, which correspond to
$p = 2.54 \times 10^{10}$ and $5.66 \times 10^{10}$ cm$^{-2}$, respectively.
In Fig.~\ref{fig:Nh13_29_of_T_+AD_noAD} the case when the alloy disorder is absent (noAD) is compared to the situation of the repulsive (+AD) alloy disorder potential.
In Fig.~\ref{fig:Nh13_45_of_T_noAD_-AD}
we show the effect of the attractive (-AD) alloy disorder potential
by referring to the data obtained without alloy disorder (noAD),
for $N_\mathrm{h} = 13$ and 45, i.e., $p = 2.54 \times 10^{10}$ and $8.78 \times 10^{10}$ cm$^{-2}$, respectively.

\begin{figure}[h!]
\includegraphics[scale=0.85]{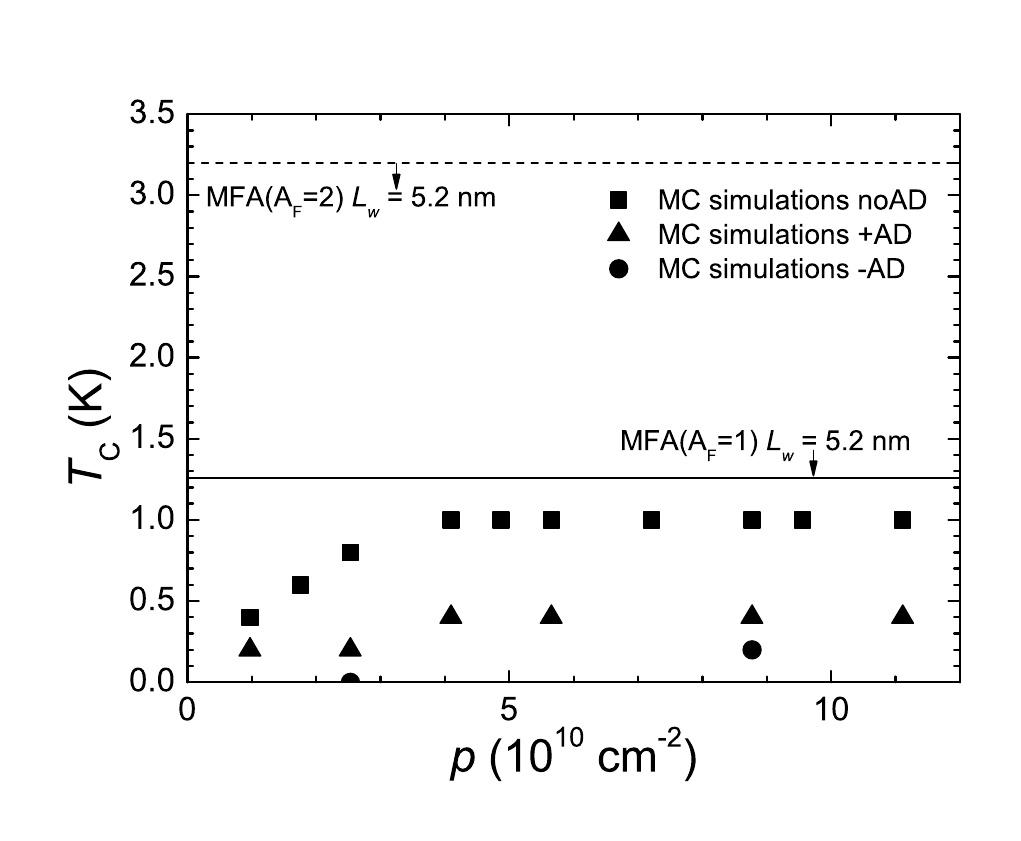}
\caption{Curie temperature $T_{\mathrm{C}}$ as a function of
the hole area density $p$.
Solid symbols show results of Monte Carlo simulations carried out for classical spins, QW width $L_{\mathrm{W}} \approx 5.2$~nm, and Mn content 4$\%$
neglecting carrier correlations in the absence of alloy disorder ($\delta V = 0$, squares)
as well as for repulsive and attractive alloy potentials ($\delta V  > 0$, triangles and $\delta V < 0$ circles, respectively).
Results obtained within the mean-field approximation (i.e., disregarding thermal
and chemical fluctuations) are shown by the solid line. The dashed line presents the MFA results taking into account carrier correlations with the Landau parameter $A_{\mathrm{F}} = 2$.}
\label{fig:TC_MC_MFA}
\end{figure}

In Fig.~\ref{fig:TC_MC_MFA}
we summarize our MC results for $T_{\mathrm{C}}$ as a function of the hole concentration obtained neglecting alloy disorder ($\delta V = 0$) as well as for $\delta V = \pm 3.93 \times 10^{-2}$~eV~nm$^3$, corresponding to the repulsive and attractive alloy potential introduced by Mn, respectively. We see that the attractive potential washes the ferromagnetism virtually entirely out. A contour plot of hole eigenfunctions demonstrates that holes get localized for the chosen magnitude of $\delta V$. Our results substantiate therefore the view that delocalized or weakly localized carriers are indispensable to set a long-range order between diluted spins.  Obviously, smaller amplitude of the attractive potential or greater values of hole concentrations will lead to carrier delocalization and the reentrance of a ferromagnetic order.

Interestingly, as seen in Fig.~\ref{fig:TC_MC_MFA}, the repulsive potential $\delta V > 0$, the case of (Cd,Mn)Te, also leads to reduced magnitudes of $T_{\mathrm{C}}$ comparing to the values determined for $\delta V = 0$. We interpret this finding by noting that in the presence of a repulsive potential, the amplitude of the wave function at Mn ions is diminished comparing to the case $\delta V = 0$. This effectively reduces the $p$-$d$ coupling and shifts the appearance of the carrier-mediated ferromagnetic order to lower temperatures.

It is instructive to compare these MC findings to the expectations of the mean-field approximation (MFA),\cite{Dietl:1997,Haury:1997,Boukari:2002,Kechrakos:2005}
derived neglecting entirely chemical and thermal fluctuations. We see in Fig.~\ref{fig:TC_MC_MFA}.
that, somewhat fortunately, the MC data for $\delta V = 0$ and MFA results for classical spins (solid line)
are in a good quantitative agreement at high hole densities.
In particular, both approaches predict that the magnitude of $T_{\mathrm{C}}$ does not vary with the hole density,
the result reflecting the energy independence of the density of states in the 2D case.
However, at low hole densities, $T_{\mathrm{C}}$ values obtained from the MC simulations tend to decrease,
the effect associated with a broadening of the density of states induced by hole scattering on Mn spins\cite{Dietl:1997,Boukari:2002} and encompassed by our MC simulations. The scattering-induced lowering of $T_{\mathrm{C}}$ is even more apparent if $\delta V > 0$.

\begin{figure}
\includegraphics[scale=0.45]{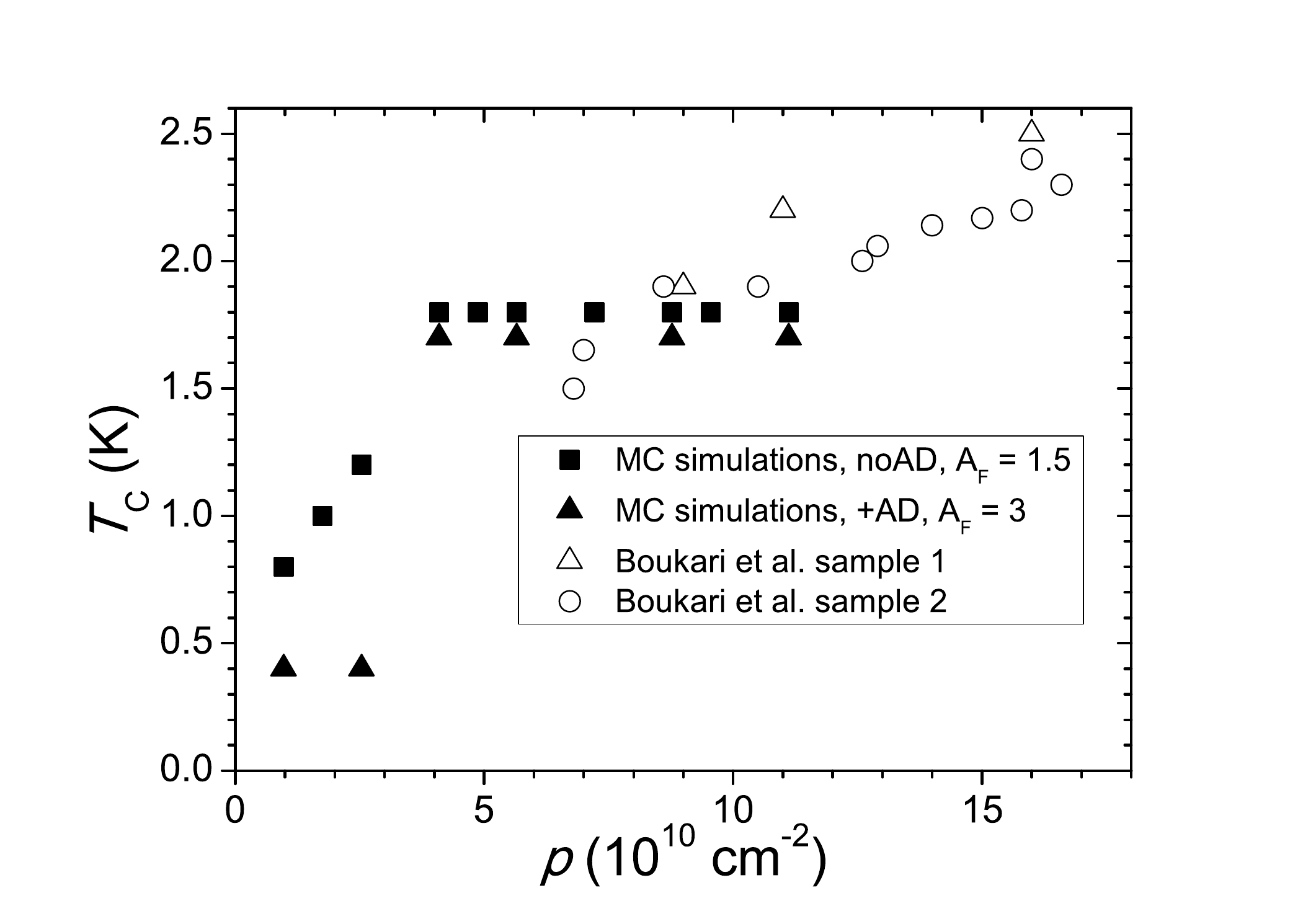}
\caption{Curie temperature $T_{\mathrm{C}}$ as a function of
the hole areal density $p$. Solid symbols show results of Monte Carlo simulations carried out for classical spins, QW width $L_{\mathrm{W}} \approx 5.2$~nm, and the Mn content 4$\%$ in the absence of alloy disorder ($\delta V = 0$, squares) and for repulsive alloy potentials ($\delta V > 0$, triangles) obtained with the Fermi liquid Landau parameters $A_F = 1.5$ and $A_F = 3$.
Experimental results\cite{Boukari:2002} for QW width $L_{\mathrm{W}} \approx $ 8~nm and Mn content 3-5$\%$ for two samples (open circles and triangles).}
\label{fig:theory_Boukari}
\end{figure}

Owing to mutual compensations between ferromagnetic and antiferromagnetic interactions, the magnitude of $T_{\mathrm{C}}$ is rather sensitive to the presence of hole correlations. This is shown by dashed line in Fig.~\ref{fig:TC_MC_MFA}, which has been obtained by incorporating into theory within the MFA  the Fermi liquid Landau parameter $A_{\mathrm{F}}  = 2$. This procedure increases the ferromagnetic contribution to $T_{\mathrm{C}}$ by a factor of two but since carrier correlations do not affect the compensating AFM term, the resulting increase of $T_{\mathrm{C}}$ is actually much greater. Therefore, in order to compare MC results to experimental findings we treat $A_{\mathrm{F}}$ as a fitting parameter. The theoretical values of  $T_{\mathrm{C}}$ displayed in Fig.~\ref{fig:theory_Boukari} has been determined from,\cite{Dietl:1997,Haury:1997,Boukari:2002,Kechrakos:2005}

\begin{equation}
\chi(T_{\mathrm{C}}) = \chi(T_{\mathrm{C}}^{(\mathrm{MC})})/A_F.
\end{equation}
Here $\chi(T) \sim T^{-\alpha}$  is the spin susceptibility of the antiferromagnetically coupled Mn spins in the absence of the holes, where $\alpha =0.79$ according to our MC simulation, and $T_{\mathrm{C}}^{(\mathrm{MC})}$ is the Curie temperature at a given hole concentration, determined from the MC simulations that neglect carrier correlations ($A_F = 1$). We see that for $p = 1 \times 10^{11}$ cm$^{-2}$
by  taking $A_F = 1.5$ and $A_F = 3$ for the case $\delta V = 0$ and $\delta V = 3.93 \times 10^{-2}$~eV~nm$^3$, respectively
we obtain a good description of the experimental findings.
In particular, a decrease of $T_{\mathrm{C}}$ values with lowering of the hole density observed experimentally\cite{Boukari:2002} and
also visible in our MC simulations,
results from scattering broadening of the density of states,\cite{Dietl:1997} particularly relevant at low Fermi energies. However, the experiment implies that this lowering extents to higher concentrations than the range predicted by the simulations, which indicates that additional scattering mechanisms operate in real samples.

Having discussed the ferromagnetic ordering temperature $T_{\mathrm{C}}$, we turn to the temperature dependence of Mn and hole spin projections, as depicted in Figs.~\ref{fig:Nh13_29_of_T_+AD_noAD} and \ref{fig:Nh13_45_of_T_noAD_-AD}. We see that the holes become entirely spin polarized, $\langle |s_z| \rangle \rightarrow 1/2$, immediately below $T_{\mathrm{C}}$. On the other hand, the increase of the Mn spin projection $\langle |S_z| \rangle$ on lowering temperature is much slower. This is because, the molecular field produced by the spin polarized carriers
is typically below 1~kOe, much too small, even at 0.1~K, to saturate Mn spins which are coupled by AFM interactions.

\begin{figure}
\includegraphics[scale=0.75]{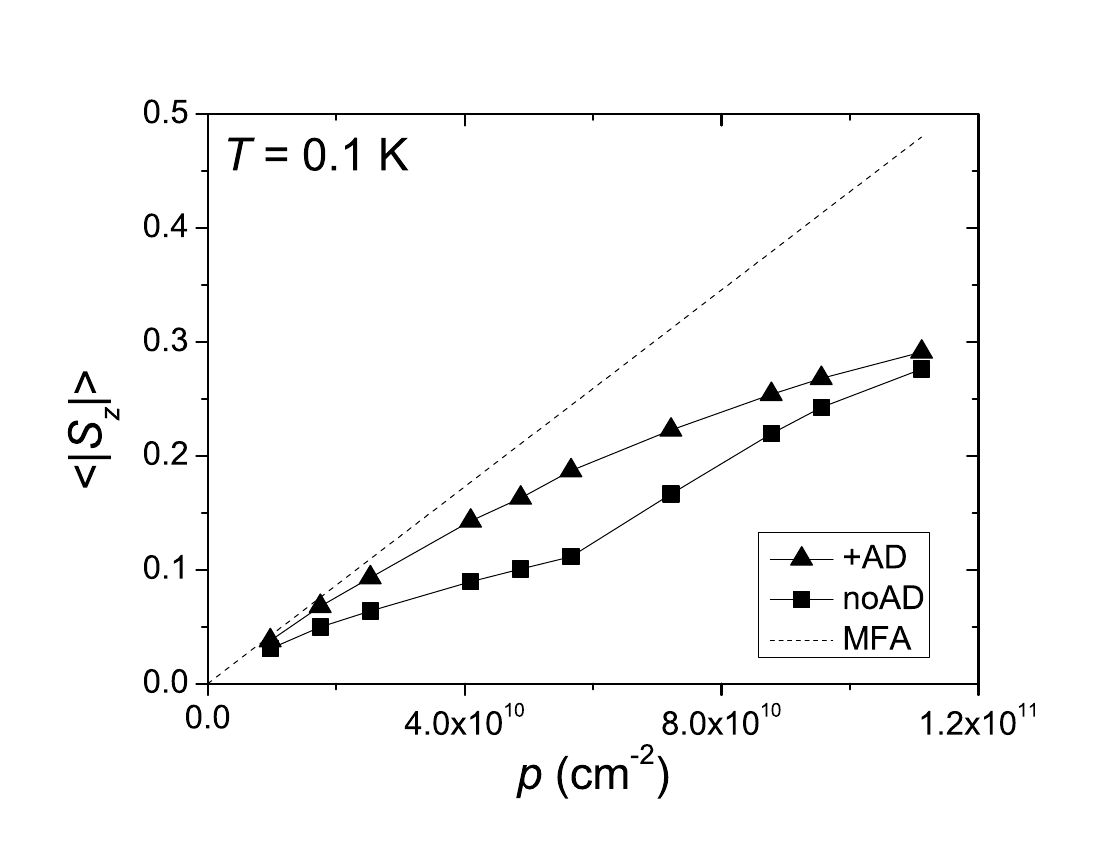}
\caption{Monte Carlo simulations:
$z$-component of the absolute spin projection of Mn spins $\langle |S_z| \rangle$ at 0.1~K
in the absence of alloy disorder ($\delta V  =  0$, squares) and
for repulsive alloy potential ($\delta V > 0$, triangles). The dashed straight line represents a dependence expected within the MFA.}
\label{fig:msiz_of_p_noAD_+AD_T=0.1K}
\end{figure}

In Fig.~\ref{fig:msiz_of_p_noAD_+AD_T=0.1K} we compare the magnitudes of the Mn spin projection $\langle |S_z| \rangle$
from our MC simulations at 0.1~K with the values expected from the MFA for the saturated carrier spins,\cite{Dietl:1997,Boukari:2002}
\begin{equation}
\langle |S_z| \rangle (T) = \frac{2^{1/2}S^2\beta p\chi(T)}{6k_{\mathrm{B}}T\chi_o(T)L_W},
\end{equation}
where $\chi(T)$ and $\chi_o(T)$ are the values of Mn spin susceptibilities per Mn ion computed without holes, in the presence and in the absence of the AF interactions, respectively. We see that the MFA quite correctly reproduces the computed magnitudes of $\langle |S_z| \rangle$ as a function of the areal hole concentration $p$.
Furthermore, while  the values of $T_{\mathrm{C}}$ are systematically smaller in the presence of the repulsive alloy potential comparing to the case $\delta V N_0 = 0$, the magnitude of $\langle |S_z| \rangle$ is seen to be larger in this case. We assign this surprising result to the fact that the repulsive alloy potential delocalizes majority-spin holes,
which can therefore embrace more effectively the magnetic impurities. However, as the hole density $p$ increases, the data with and without alloy potential tend to approach. This is because the dense majority-spin holes can embrace more effectively the magnetic impurities even without help from the repulsive potential.  In the other extreme, when the values of $p$ decreases towards zero, the data with and without alloy potential approach, too. Here, the $T_{\mathrm{C}}$ magnitude becomes so low that the carriers' spins ceases to be entirely saturated in the presence of the repulsive potential, which lowers the corresponding $\langle |S_z| \rangle$ value.

\section{Interpretation of magneto-optical measurements}
\label{sec:dyn-dom}
In this section we present our MC results on Mn magnetization dynamics as a function of
temperature and carrier concentration.
We also discuss whether it is possible within the proposed model to estimate the typical size
of magnetic domains.

\subsection{Magnetization dynamics: autocorrelation function}
\label{subsec:dyn}
\begin{figure}
\includegraphics[scale=1.0]{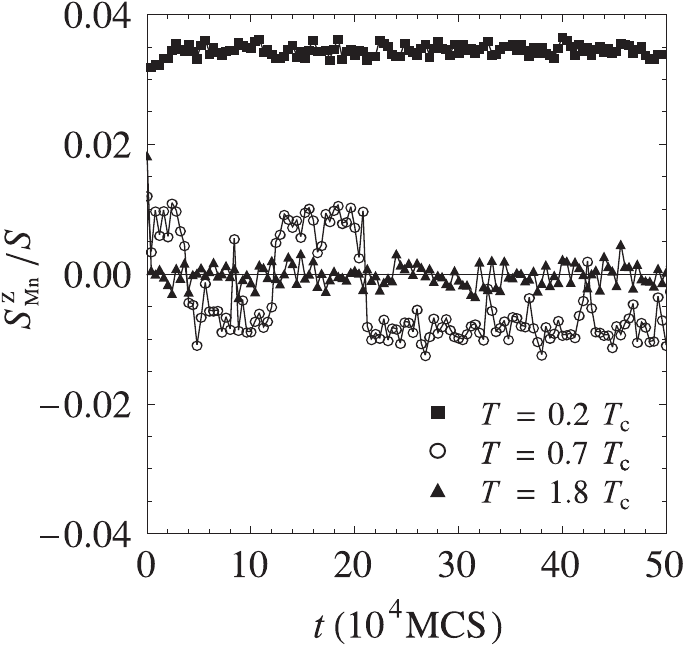}
\caption{Monte Carlo simulations: Time evolution of  $S^{z}_{Mn}(t)/S$  at different temperatures  in a
Cd$_{0.96}$Mn$_{0.04}$Te quantum well. The hole concentration is
$p = 0.4 \times 10^{11} \ cm^{-2}$ ($N_{h} = 21$). Time is
measured in the Monte Carlo steps per Mn site.} \label{fig:corfun9}
\end{figure}

We have traced time dependence of Mn magnetization in the absence of an external magnetic field for very long runs up to $10^5$ MC sweeps. We have repeated our simulations for different initial configurations including the one with all Mn spins aligned along the $z$-axis. From our findings we conclude that except for the first few MC sweeps the whole time evolution of magnetization does not depend on the initial configuration. The typical time-evolution of magnetization at various temperatures is shown in Fig.~\ref{fig:corfun9}. For  temperatures well below $T_{\mathrm{C}}$ the system remains almost unchanged during the whole simulation time. For higher temperatures one can observe occasional global spin inversions: the system flips over between two states of equal energy and spontaneous magnetization that differ only in sign.

The typical time scale of the observed evolution depends on many factors including temperature and carrier concentration.  To determine this characteristic time scale we calculate the time-autocorrelation $G(\tau)$ of  magnetization given by\cite{Newman:1999}

\begin{equation}\label{autocor}
  G(\tau) = \frac{\left\langle m(t)m(t+\tau)\right\rangle-\left\langle m\right\rangle^{2}}{\left\langle m^{2}\right\rangle-\left\langle m\right\rangle^{2}},
\end{equation}

where $m(t)$ is the value of  magnetization at time $t$ and
$\left\langle m\right\rangle$ is its average value. The autocorrelation function defined by  Eq.~\ref{autocor} depends only on the time difference $\tau$. To calculate this autocorrelation function for up to $\tau=10^{4}$ MCS computation runs as long as at least $10^{5}$ MCS are needed. The total computer time depends on numerical
parameters of which the most important is the $k$-space cut-off.

\begin{figure}
\includegraphics[scale=1.0]{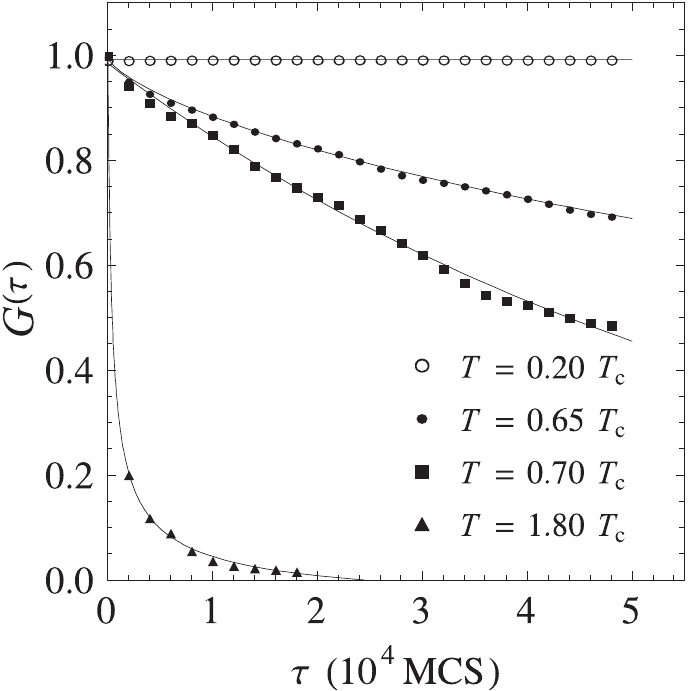}
\caption{Monte Carlo simulations: The magnetization autocorrelation function $G(\tau)$
of Mn ions in a Cd$_{0.96}$Mn$_{0.04}$Te quantum well calculated at different temperatures.
The concentration of holes $p = 0.4 \times 10^{11}$~cm$^{-2}$ ($N_{h} = 21$). Time is
measured in the Monte Carlo steps per one Mn site. The thin solid lines describe  exponential
decays fitted to the Monte Carlo results} \label{fig:corfun3}
\end{figure}

\begin{figure}
\includegraphics[scale=1.0]{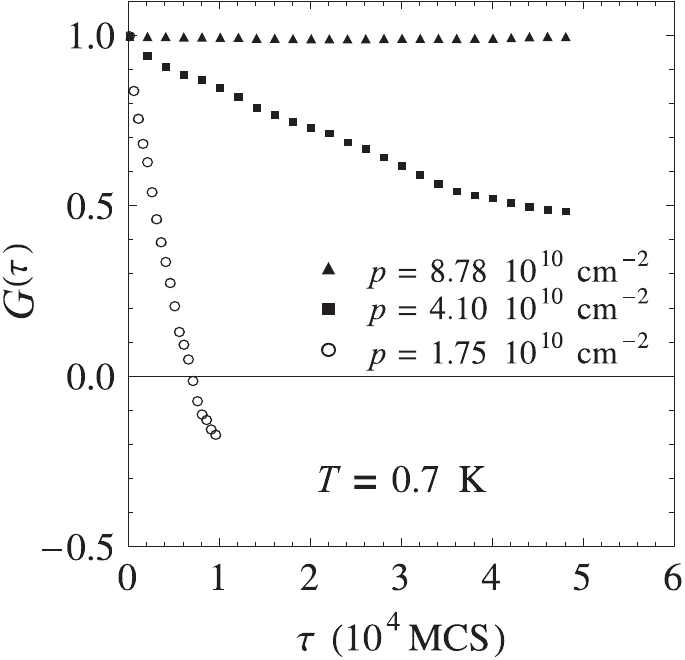}
\caption{Monte Carlo simulations: The magnetization autocorrelation function $G(\tau)$
for selected values of  hole concentrations  in a
Cd$_{0.96}$Mn$_{0.04}$Te quantum well at $T = 0.7$~K. Time is
measured in the Monte Carlo steps per Mn site.} \label{fig:corfun4}
\end{figure}

Figure~\ref{fig:corfun3} shows the magnetization autocorrelation functions calculated for the studied system
at various temperatures below and above  $T_{\mathrm{C}}$. Above $T_{\mathrm{C}}$, i.e., in
the paramagnetic state, one can observe a very fast decay of the autocorrelation function whereas for temperatures
below $T_{\mathrm{C}}$ correlations persist for times orders of magnitude longer. Within the present MC approach, it is not possible to translate numbers of MC steps into physical time units in a reliable quantitative way. The qualitative relations between the time scales of the investigated magnetic relaxation processes should nevertheless be reproduced in a correct way. Therefore, we undertake only a qualitative comparison with the experimental data of Sec.~\ref{sec:experiments}.

In order to better understand mechanisms that account for magnetization dynamics in a system of competing spin-spin AFM and carrier-mediated ferromagnetic interactions we have also calculated autocorrelation functions for a variety of carrier concentration values keeping the number of Mn ions unchanged. In Fig.~\ref{fig:corfun4} we display the corresponding results calculated for three different carrier concentrations at the same temperature  $T = 0.7$~K. As one could expect we do observe a significant slowing down of magnetization dynamics with increasing the number of holes in the QW. Such behavior is understandable within the proposed model. Taking into account the magnetic anisotropy -- all hole-spins align along $z$-axis -- and the fact that only the $z$-component of each Mn-spin couples via the $p$-$d$ exchange to the system of holes, one can see that if more holes are present in the system, the value of local magnetization and, hence, the barrier height for magnetization reversal become higher.
Note that for the carrier concentrations $p = 4.10 \cdot 10^{10}$~cm$^{-2}$ and $p = 8.78 \cdot 10^{10}$~cm$^{-2}$ the system is in the ferromagnetic state at $T = 0.7$~K. In contrast,  for $p = 1.75 \cdot 10^{10}$~cm$^{-2}$ the Curie temperature lies below $T = 0.7$~K (cf. next sections), so that the corresponding points describe the dynamics of magnetization in a paramagnetic QW.

\begin{figure}
\includegraphics[scale=1.0]{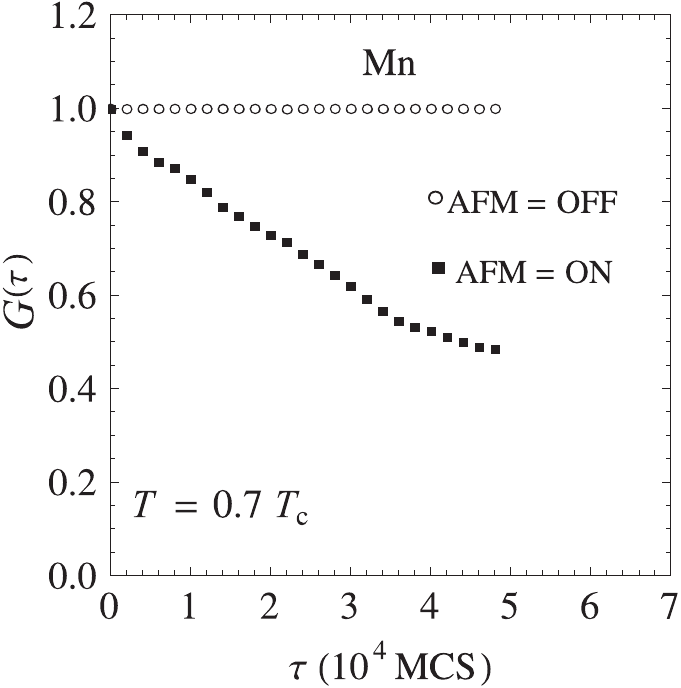}
\caption{Monte Carlo simulations: The magnetization autocorrelation function $G(\tau)$ of Mn ions in a Cd$_{0.96}$Mn$_{0.04}$Te quantum well at $T = 0.7 T_{\mathrm{C}}$. The hole density is $p = 0.4 \times 10^{11}$~cm$^{-2}$ ($N_{h} = 21$). Open circles and full squares:  antiferromagnetic interactions are neglected ($J_{ij} = 0$) and taken into account ($J_{ij} \neq 0$), respectively. Time is measured in the Monte Carlo steps per Mn site.} \label{fig:corfun5}
\end{figure}

\begin{figure}
\includegraphics[scale=1.0]{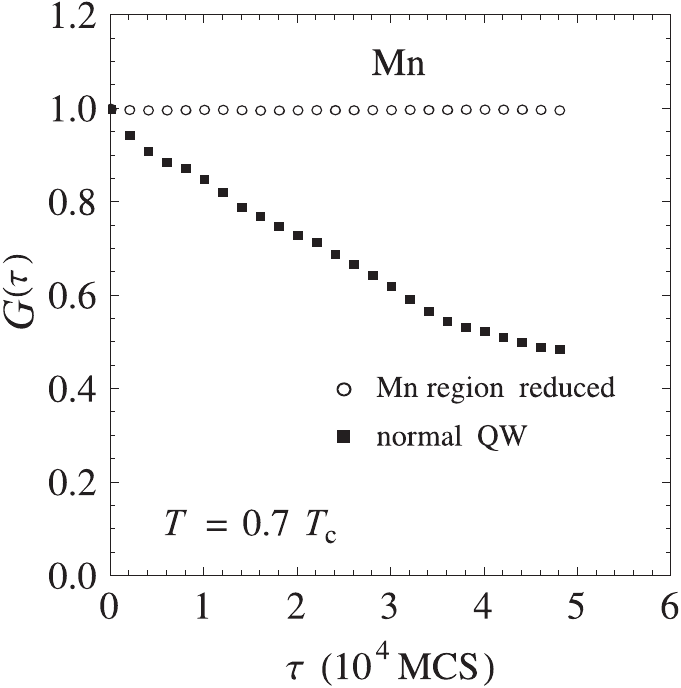}
\caption{Monte Carlo simulations: The magnetization autocorrelation function $G(\tau)$ of Mn ions in a Cd$_{0.96}$Mn$_{0.04}$Te quantum well at $T = 0.7 T_{\mathrm{C}}$. The hole density is $p = 0.4 \times 10^{11}$~cm$^{-2}$ ($N_{h} = 21$).Open circles and full squares: Mn located in the central layers of the
QW only and evenly distributed within the whole QW, respectively. Time is measured in the Monte Carlo steps
per Mn site.} \label{fig:corfun11}
\end{figure}

A next important aspect one must understand in order to interpret properly the experimental findings is the way in which the AFM interactions influence the magnetization dynamics. Therefore, we have calculated the
magnetization autocorrelation function for the system where all short-range AFM interactions are disregarded.
In Fig.~\ref{fig:corfun5} we compare the obtained magnetic autocorrelation functions for systems with and without AFM interactions calculated at $T=0.7T_{\mathrm{C}}$ (for each system its own $T_{\mathrm{C}}$ as previously determined by MC simulations\cite{Kechrakos:2005} had been used). The results presented in Fig.~\ref{fig:corfun5} clearly indicate that short-range AFM interactions may strongly accelerate the decay of the ferromagnetic order. In accord with the above observation one can expect that the effect of AFM interactions is
considerably reduced when the layer of Mn ions is thinner than the extend of the hole wave function, in other words, when Mn ions are concentrated close to the maximum (probability) density of holes. To check this we repeated our calculations with all Mn ions occupying only the four central layers of the investigated QW (note that it is eight layers' thick and that for holes the ground state of the infinite well is taken as the "perpendicular" part of their wave function, $\varphi(z)$). The results of our calculations are shown in Fig.~\ref{fig:corfun11}. One can clearly see that indeed if Mn ions reside only in the central layer, the magnetization dynamics resembles the one for the case when all AFM interactions are switched off. We may as well formulate our conclusion the other way round: more
Mn spins close to the QW edges,  faster  relaxation of magnetization. This finding provides a support for theoretical considerations,\cite{Boudinet:1993}
suggesting that magnetization relaxation of bound magnetic polarons in $p$-type CdTe/(Cd,Mn)Te QWs occurs
owing to the AFM coupling to the Mn spins located outside the relevant Bohr radius.

\subsection{Measuring the domain size: the two-point spin-spin connected correlation function}
\label{subsec:dom}
Let us now look at the spatial distribution of magnetization and, in particular, whether the presence of competing short-range antiferromagnetic and long-range ferromagnetic interactions results in the formation of magnetic domains.
One of the standard ways to estimate the typical domain size is to calculate the spin-spin two-point
connected correlation function $G_{c}^{2}(R)$ given by\cite{Newman:1999}

\begin{equation}
  G_{c}^{2}(R) = \frac{\left\langle m(\textbf{r})m(\textbf{r}+\textbf{R})\right\rangle-\left\langle m\right\rangle^{2}}{\left\langle m^{2}\right\rangle-\left\langle m\right\rangle^{2}},
\end{equation}

\begin{figure}
\includegraphics[scale=1.0]{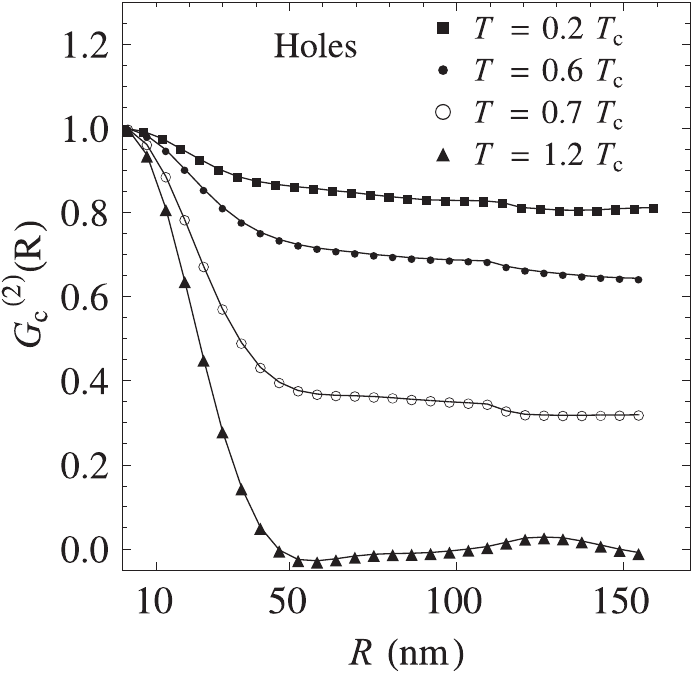}
\caption{Monte Carlo simulations: The two-point connected correlation function $G^{(2)}_{c}(R)$ of valence band holes at various temperatures  in a  Cd$_{0.96}$Mn$_{0.04}$Te quantum well. The hole concentration is $p = 0.4 \times 10^{11}$~cm$^{-2}$ ($N_{h} = 21$).} \label{fig:corfun6}
\end{figure}

\begin{figure}
\includegraphics[scale=1.0]{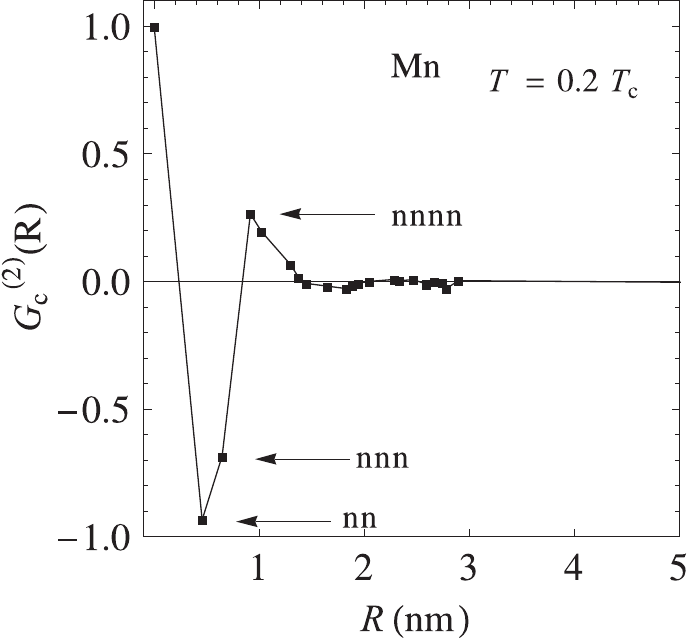}
\caption{Monte Carlo simulations: The two-point connected correlation function $G^{(2)}_{c}(R)$ of Mn ions in a Cd$_{0.96}$Mn$_{0.04}$Te quantum well at $T=0.2T_{\mathrm{C}}$. The hole concentration is $p = 0.4 \times 10^{11}$~cm$^{-2}$ ($N_{h} = 21$).} \label{fig:corfun7}
\end{figure}

\begin{figure}
\includegraphics[scale=1.0]{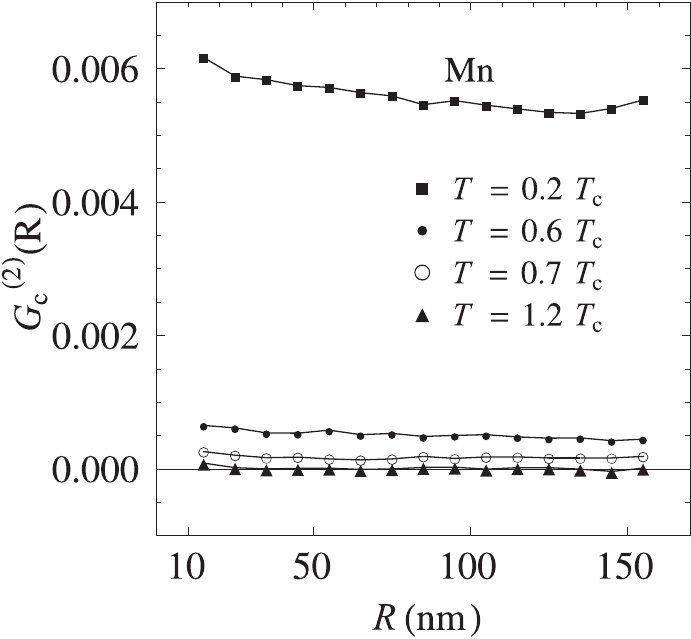}
\caption{Monte Carlo Simulations: The long-distance parts of two-point connected correlation functions $G^{(2)}_{c}(R)$ of Mn ions at various temperatures  in a Cd$_{0.96}$Mn$_{0.04}$Te quantum well. The hole concentration is $p = 0.4 \times 10^{11}$~cm$^{-2}$ ($N_{h} = 21$).} \label{fig:corfun8}
\end{figure}

We have calculated spin-spin correlation functions for both: the localized Mn ions and the holes for a number of temperature values and various carrier concentrations. Figure~\ref{fig:corfun6} shows the correlation functions $G_{c}^{2}(R)$ obtained for the hole subsystem. We observe long-range  correlations (longer than the size of the simulation box) in the ferromagnetic state. For temperatures  below $T_{\mathrm{C}}$ the hole liquid becomes completely spin-polarized. This has already been predicted in the frame of the mean-field approximation for the 2D hole gas in a QW.\cite{Dietl:1997} In Fig.~\ref{fig:corfun7} and \ref{fig:corfun8} the spin-spin correlation functions for the Mn  subsystem are shown. In Fig.~\ref{fig:corfun7} the scale has been stretched to make visible how the short-range AFM interactions influence the spin-spin correlation function at small distances. In this way we can clearly see that for the nearest neighbor Mn spins the AFM coupling prevails (the correlation function becomes negative). At long distances, however, below  $T_{\mathrm{C}}$, the long-range ferromagnetic interactions dominate as may be seen in  Fig.~\ref{fig:corfun8} that shows the spin-spin correlation functions on a larger scale. As the distance $R$ between the Mn spins becomes large we observe (small) positive long-range correlations for temperatures below $T_{\mathrm{C}}$. At the same time the correlations decrease to zero when temperature increases towards $T_{\mathrm{C}}$. The positive correlations observed in the ferromagnetic phase and the lack of correlations in the paramagnetic phase may indicate that magnetic domains do exist, but their typical size is greater than the simulation box. To verify this conjecture further massive simulations would be needed.

\section{Conclusions}
\label{sec:conclusion}
We have examined magnetization dynamics in a p-Cd$_{0.96}$Mn$_{0.04}$Te quantum well by magneto-optical studies and by Monte Carlo simulations taking into account the presence of competing short-range antiferromagnetic superexchange and long-range   carrier-mediated ferromagnetic interactions. Single-particle hole energies and eigenfunctions for particular Mn spin configurations have been updated numerically after each Monte Carlo sweep. In addition to the $p$-$d$ exchange interaction, spin-independent alloy disorder has been taken into account and found to be important. In particular, a sufficiently strong attractive potential introduced by the magnetic ion leads to hole localization and to the corresponding disappearance of a long-range order mediated by the carriers. Also a strong repulsive potential, by reducing the magnitude of the carrier wave function at  the magnetic ion, may diminish the magnitude of $T_{\mathrm{C}}$. This effect is particularly strong in the case of a quantum well, where the direction of the hole spin is fixed rather by the spin-orbit interaction than by the $p$-$d$ coupling, so that the repulsive potential may not be compensated by the $p$-$d$ interaction.

Both experimental
and theoretical results demonstrate that the transition from the paramagnetic to ferromagnetic phase
increases the magnetization relaxation time. In particular, the relaxation time determined from tracing the time evolution of magnetization after a pulse of the magnetic field is shorter than 20 ns in the paramagnetic state. Lowering temperature below $T_{\mathrm{C}}$ results in
an increase of the relaxation time up to only 2 $\mu$s. Our numerical results  indicate
that antiferromagnetic interactions between Mn spins account for this relatively fast magnetization fluctuations and,
thus, might be responsible for the absence of spontaneous magnetization below $T_{\mathrm{C}}$, as found in static measurements.
While our Monte Carlo simulations  of the two-point spin-spin connected correlation function show a huge
influence of the short-range antiferromagnetic interactions on the magnetization dynamics, they do not reveal the formation of
domains that would be smaller than our simulation cell, $L = 226$~nm.

At the same time, our Monte Carlo results reveal that magnetization relaxation in the ferromagnetic phase proceeds primarily due to antiferromagnetic couplings to Mn spins residing at the quantum well boundary, as they are weakly polarized by the carriers. Accordingly, magnetization dynamics would be much slow down if the width of the region containing Mn spins were been narrower than the extent of the carrier wave function. We predict, therefore, that spontaneous magnetization and the associated magnetic hysteresis could be observed in quantum wells containing the Mn layer only in its center.

\begin{acknowledgements}
The work in Warsaw was supported in part by Technology Agency, by SPINTRA Project of European Science Foundation (ERAS-CT-2003-980409), and by the FunDMS Advanced Grant within the European Research Council "Ideas" Programme of EC 7FP.
Part of the necessary computer time in Athens was provided by
the National Grid Infrastructure {\it HellasGrid}.
We thank N.~Papanikolaou for useful discussions.
\end{acknowledgements}

\end{document}